# Non-relativistic spin splitting in a triangular metal-excess magnet $Fe_{1+\delta}Sb$

[1]Chao-Chun Wei, [2]Xiaojuan Ni, [1,3]Sophia Adams, [4]Jacob Kjeldahl Jensen, [5]Jue Liu, [5]Qiang Zhang, [4]Luisa Whittaker-Brooks, [1]Huiwen Ji*

[1]*Department of Materials Science and Engineering, University of Utah, Salt Lake City, Utah 84112, United States*

[2]*Department of Chemistry and Biochemistry, The University of Arizona, Tucson, Arizona 85721-0041, United States*

[3]*Department of Chemical and Biomolecular Engineering, Johns Hopkins University, Baltimore, Maryland, 21218, United States*

[4]*Department of Chemistry, University of Utah, Salt Lake City, Utah 84112, United States*

[5]*Neutron Sciences Division, Oak Ridge National Laboratory, Oak Ridge, Tennessee 37831, United States*

Corresponding: huiwen.ji@utah.edu

**ABSTRACT**

Non-relativistic spin-splitting (NRSS) antiferromagnets have recently emerged as an important class of magnetic materials that combine compensated magnetism with momentum-dependent spin splitting, offering new opportunities for spintronic applications. Here, we investigate a NiAs-type $Fe_{1+\delta}Sb$ series ($\delta$ = 0.17-0.30) using neutron diffraction, pair distribution function, magnetometry and density functional theory calculations. Neutron diffraction establishes that $Fe_{1+\delta}Sb$ adopts a 120° coplanar compensated magnetic order with a non-zero propagation vector $\boldsymbol{k}$ = (1/3, 1/3, 0). Increasing interstitial Fe suppresses the ordered magnetic moment while inducing local symmetry lowering, as revealed by pair distribution function refinements. Density functional theory predicts momentum-dependent spin splitting, dominated by an out-of-plane spin polarization with an odd-parity *f*-wave-like symmetry, establishing the material as a non-collinear NRSS antiferromagnet. Motivated by the structural similarities between $Fe_{1+\delta}Sb$ and a known altermagnet CrSb, we further investigate their solid solution and find that Cr substitution at intermediate concentrations gives rise to a ferromagnetic component and a cluster spin-glass behavior. These results establish $Fe_{1+\delta}Sb$ as a new platform for non-collinear NRSS antiferromagnetism and demonstrate metal interstitial and substitution as effective parameters for tuning the magnetic order and properties.

## INTRODUCTION

Antiferromagnets (AFMs) have recently emerged as a promising platform for spintronic devices for their negligible stray fields, ultrafast spin dynamics, and robustness against external magnetic fields. While conventional AFMs possess spin-degenerate electronic bands, a special class broadly referred to as non-relativistic spin-splitting (NRSS) AFMs has attracted considerable theoretical and experimental interest because they exhibit momentum-dependent spin splitting despite having nearly zero net magnetization.[1,2] Their opposite-spin sublattices are not related by an inversion or translation symmetry.[3] Consequently, NRSS AFMs possess ferromagnetic-like spin-splitting while retaining the intrinsic advantages of AFMs. This unique combination gives rise to a plethora of unconventional physical properties, e.g., spin-polarized electric currents, spin currents, spin-splitting torque, giant magnetoresistance, and anomalous Hall and Nernst effects.[4] [5] [6] [7]

Among NRSS AFMS, those with collinear magnetic structures and rotation symmetry between the opposite-spin sublattices are termed altermagnets (AMs),[1] with notable candidates like $RuO_2$[8], MnTe[9], CrSb[10], $Fe_{1/4}NbS_2$[11], and $La_2O_3Mn_2Se_2$[12]. Recent theoretical studies have pointed out that NRSS can also arise in non-collinear AFMs so long as $PT$ and $\tau T$ are both broken, where $P$, $T$ and $\tau$ represent inversion, time-reversal, and translation symmetries, respectively.[13,14] This theoretical finding broadens the design space for exploring unique electronic and spintronic properties. For example, a Landau-theory and spin-space-group analysis on a chiral non-collinear AFM $Mn_3IrSi$ predict intricate hedgehog- and quadrupolar-like spin textures, which might lead to large spin Hall and Edelstein effects in the absence of spin-orbit coupling (SOC).[15] Recently, our group explored a high-temperature polymorph of $Cr_7Se_8$, which adopts a metal-deficient NiAs-type structure and a 120$^\circ$ coplanar magnetic order. Electronic structure calculation predicts odd-parity *f*-wave-like spin splitting and remarkable mirror-symmetry-protected Weyl nodal loops near the Fermi level, thereby combining crystal-symmetry-driven spin splitting and topologically non-trivial electronic states in one material.[16] These findings underscore the importance of discovering new non-collinear NRSS material platforms for unlocking novel properties.

Inspired by the NiAs-type structure which hosts multiple known altermagnets, $Fe_{1+\delta}Sb$ caught our attention as a promising platform for exploring non-collinear NRSS. $Fe_{1+\delta}Sb$ crystallizes in the hexagonal NiAs structure with the excess Fe occupying the interstitial 2*d* site.[17] Similar excess metal occupation in interstitial sites has also been observed for altermagnets $Cr_{1+\delta}Sb$ and believed to enhance structural stability by increased orbital hybridization.[18] Unlike the collinear magnetic order in CrSb, prior neutron diffraction and Mossbauer studies have established that $Fe_{1+\delta}Sb$ adopts a 120$^\circ$ coplanar order for the regular-lattice Fe moments, while the interstitial Fe moments remain paramagnetic below the Néel temperature $T_N$ [17,19] and form local magnetic clusters that freeze at low temperatures.[20] Meanwhile, doping provides an additional tuning parameter. Early studies have demonstrated that $(Fe,Cr)_{1+\delta}Sb$ and $(Fe,Co)_{1+\delta}Sb$ form continuous solid solutions and exhibit a systematic evolution of the magnetic ground state with composition, indicative of competition among Fe-Fe, Fe-Cr/Co, and Cr/Co-Cr/Co exchange interactions.

In this work, we combine neutron powder diffraction, magnetic property measurements, and first-principles calculations to revisit $Fe_{1+\delta}Sb$ as well as its Cr-doped series. We first establish $Fe_{1+\delta}Sb$, regardless of the excess Fe concentration, as a non-collinear NRSS AFM with a compensated 120° triangular coplanar magnetic structure, based on the neutron diffraction. Density functional theory calculations predict an *f*-wave-like ***k***-dependent spin polarization in the absence of SOC with a predominant $S_z$ component despite the in-plane moments. We then show that Cr substitution gives rise to enhanced low-temperature magnetization and glassy dynamics at intermediate Cr concentrations. These results identify a new non-collinear NRSS platform featuring metal excess and doping as viable tuning parameters.

## METHODS

All $(Fe, Cr)_{1+\delta}Sb$ samples were synthesized by mixing stoichiometric Fe powder (Sigma Aldrich, 99.9%), Cr powder (Sigma Aldrich, 99%), and Sb metal chunk (Luciteria Science, 99.99%). The precursors were ground, pelletized, and sealed in a quartz ampule. The sample was first heated to 650°C for 48 h, followed by regrinding, and a second heat treatment at 950°C for 48 h, followed by water quenching.

DC magnetization measurements ($M$ vs. $T$ and $M$ vs. $H$) were measured using a Quantum Design Physical Properties Measurement System (PPMS) with a vibrating-sample magnetometer (VSM) option. Susceptibility was obtained by normalizing the measured magnetization by the field. AC susceptibility experiments were carried out using an ACMS option with a 10-Oe excitation field at 100-10000 Hz. Powder neutron diffraction was measured at POWGEN and NOMAD at SNS, ORNL. Approximately 1.3 g powder was loaded into a 6-mm diameter vanadium PAC can, which was subsequently sealed with helium exchange gas to ensure optimal thermal conductivity. Data were collected at the designated temperature using a single neutron frame centered at a wavelength of 1.5 Å, providing a $d$-spacing coverage of 0.5–12 Å. All the data were reduced by subtracting the empty PAC container signal, with normalization performed using the difference between the vanadium standard and the empty instrument background. Neutron diffraction data was refined using the GSAS-II package[21], and the pair distribution function (PDF) data was refined using PDFgui[22].

A STOE STADI P x-ray diffractometer was used to collect powder diffraction data in transmission geometry with a Mo anode with K$\alpha$1 radiation ($\lambda = 0.7094$ Å), a Ge (111) curved monochromator, and a Dectris MYTHEN2 detector. Data was collected at room temperature in static mode across the $2\theta$ angle range of 2° to 60°. Structural refinement was done with the GSAS-II software package. VESTA was used to visualize crystal structures.

Density functional theory (DFT) calculations for FeSb were performed within the generalized gradient approximation (GGA) using the Perdew–Burke–Ernzerhof (PBE) exchange–correlation functional, as implemented in the Vienna Ab initio Simulation Package (VASP).[23-25] The Brillouin zone was sampled using a 7 × 7 × 11 Monkhorst–Pack k-point mesh, and the plane-wave kinetic-energy cutoff was set to 500 eV. All atomic positions were fully relaxed until the residual forces on each atom were smaller than 0.01 eV/Å. Since the Fe 3*d* electrons exhibit strong electronic correlations, on-site Coulomb interactions were treated using the DFT+U approach.[26] The on-site Coulomb and exchange interaction parameters were set to $U$ = 5 eV and $J$ = 1 eV, respectively, as these values have been shown to provide a reliable description of the Fe 3d electronic states.[27] The experimentally observed coplanar magnetic configuration was investigated using noncollinear calculations.[28] Spin–orbit coupling (SOC) was included in the electronic-structure calculations. The calculated results were subsequently analyzed and post-processed using the VASPKIT package.[29]

## RESULTS AND DISCUSSION

Fig. 1a shows the hexagonal NiAs-type crystal structure of $Fe_{1+\delta}Sb$. Over-stoichiometric Fe atoms occupy the interstitial 2*d* Wyckoff site, which has the same site symmetry ($\bar{6}m2$) as Sb atoms. The purity of solid-state synthesized $Fe_{1+\delta}Sb$ with $\delta$= 0.2, 0.3, and 0.4 is confirmed by neutron powder diffraction, shown in Fig. 1b. The Rietveld refinement using a NiAs-type structural model yields excellent agreement with the observations. The corresponding refinement parameters are summarized in Table S1. Increasing Fe excess leads to an expanded lattice in both $a$ and $c$. The refined interstitial Fe site occupancy is 0.1748(13), 0.2295(18), and 0.3018(19) for the starting compositions of $Fe_{1.2}Sb$, $Fe_{1.3}Sb$, and $Fe_{1.4}Sb$, respectively. The discrepancies might be due to Fe powder loss during initial handling and mixing. We thus refer to them as $Fe_{1.17}Sb$, $Fe_{1.23}Sb$, and $Fe_{1.30}Sb$ in the remaining text.

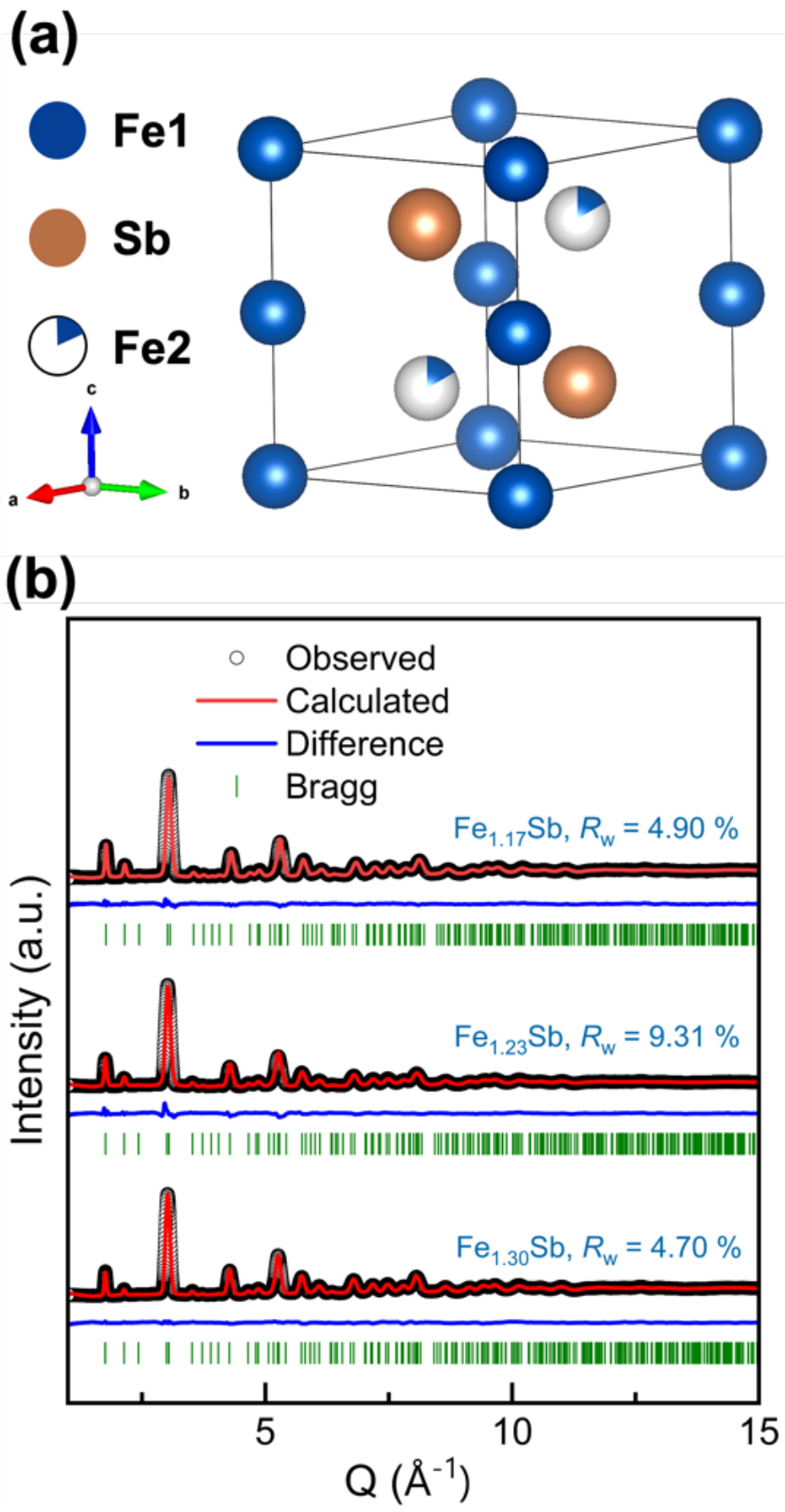

**Figure 1.** (a) Crystal Structure of $Fe_{1+\delta}Sb$ in a NiAs-type framework. Fe1 atoms occupy the regular NiAs lattice sites, while Fe2 denotes the excess Fe atoms occupying the interstitial *2d* Wyckoff sites. (b) Rietveld refinement of neutron powder diffraction of $Fe_{1.17}Sb$, $Fe_{1.23}Sb$, and $Fe_{1.30}Sb$. Black circles, red and blue curves, and olive ticks represent observation, calculation, fit residual, and Bragg positions, respectively.

To examine any subtle structural impact of interstitial Fe, we further investigate the local atomic structures by refining the neutron pair distribution function (NPDF) data, as shown in Fig 2. As summarized in Table S2 and in Fig 2a, the average NiAs-type structural model provides good fit for $Fe_{1.17}Sb$ and $Fe_{1.23}Sb$, yielding $R_w = 7.08\%$, while the refinement quality noticeably deteriorates for $Fe_{1.30}Sb$, with $R_w$ approaching 10%, suggesting growing short-range order that deviates from the average hexagonal structure with the interstitial Fe. To account for this deviation, we explore lower-symmetry models and find that the orthorhombic *Amm*2 model, as illustrated in Fig 2c, provides a significantly improved description of the local structure. The transformation from the NiAs-type unit cell to the *Amm*2 setting is given by:

$$(a_{Amm2}, b_{Amm2}, c_{Amm2)} = (a_{NiAs}, b_{NiAs}, c_{NiAs}) \begin{pmatrix} 0 & 1 & 1 \\ 0 & 1 & -1 \\ -1 & 0 & 0 \end{pmatrix} + \begin{pmatrix} 0 \\ 0 \\ 0.25 \end{pmatrix}$$

where the matrix presents the rotational transformation, and the translation vector shifts the origin. As summarized in Table S3 and shown in Fig 2b, the *Amm2* structure improves the fit across all compositions and particularly for $Fe_{1.30}Sb$, implying that increasing interstitial Fe induces more local symmetry lowering. The refined model suggests that interstitial Fe occupies the Fe2 site more favorably than the Fe3 site, resulting in an alternating-layer arrangement. The interstitial Fe order is further accompanied by in-plane displacement of the Sb2 atoms in the same layer as the Fe3 site. A similar orthorhombic structure was recently reported in MnTe from the same structural family, with atomic-resolution scanning transmission electron microscopy (STEM) revealing similar local inversion-symmetry-breaking.[30] Their group analysis shows that the local distortion originate from $\Gamma$-point distortion modes that locally condense into $Amm2(\Gamma_4^- \otimes \Gamma_6^-)$ and $Cmc2(\Gamma_2^- \otimes \Gamma_5^-)$. These structural distortions precede magnetic ordering above $T_N$. In our case, the correlated atomic displacements (e.g., for Sb2) associated with the orthorhombic distortion might be induced by additional in-plane Fe-vacancy or charge order that breaks the 6-fold rotation symmetry, although they cannot be reliably resolved without more advanced techniques such as single-crystal diffuse scattering or symmetry-sensitive Raman spectroscopy.

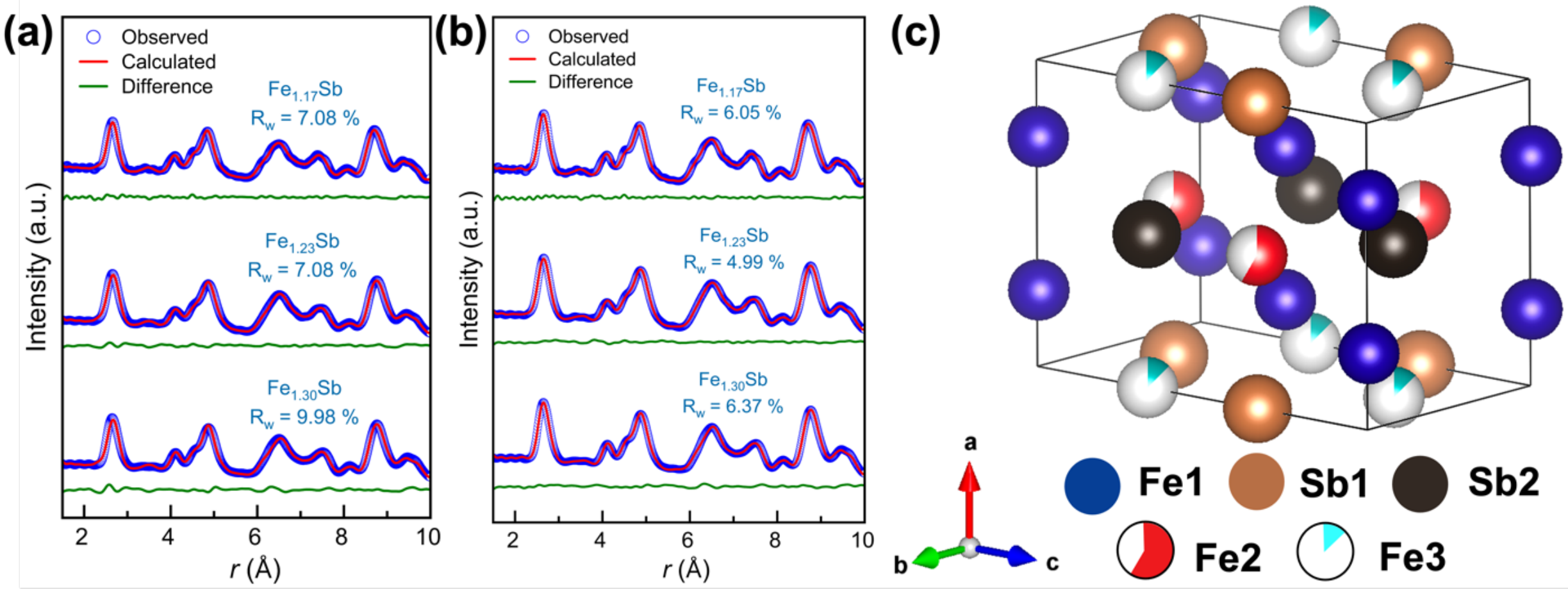

**Figure 2.** NPDF refinements of $Fe_{1+\delta}Sb$ over the real-space range of 1.5 – 10 Å using (a) the hexagonal NiAs-type ($P6_3/mmc$) structural model and (b) orthorhombic *Amm*2 structural model, and the refined weighted residual $R_w$ are indicated for each composition. (c) Crystal structure of *Amm*2 model.

The magnetic properties of $Fe_{1+\delta}Sb$ are investigated by DC magnetization under zero-field cooling (ZFC) and field cooling (FC) with an applied field of 500 Oe over the temperature range between 2 and 300 K, as shown in Fig. 3. In $Fe_{1.17}Sb$, a broad maximum appears at ~20 K, below which a bifurcation between the ZFC and FC curves is observed. In addition, a faint kink is observed at ~ 170 K (enlarged view in Fig. 3a). The absence of a clear magnetic anomaly that indicates the Néel temperatures ($T_N$'s) is due to the large paramagnetic contribution from interstitial Fe atoms. [17,19] The magnetization maxima and the accompanying bifurcation shift higher to ~50 K and ~80 K in $Fe_{1.23}Sb$ and $Fe_{1.30}Sb$, respectively. Meanwhile, the small AFM-like kink in $Fe_{1.23}Sb$ occurs at ~

110 K and becomes invisible in $Fe_{1.30}Sb$ likely due to the strongest paramagnetic background in the latter. Since previous studies based on magnetic properties, neutron diffraction, and Mössbauer spectroscopy have indicated that $T_N$ falls between 100 and 220 K depending on the interstitial Fe concentration, the weak anomalies above 100 K we observe in magnetization might be associated with AFM transitions. The lower-temperature maximum, on the hand, has been associated with the freezing of magnetic clusters formed by the interstitial Fe moments, which constitutes a spin-glass transition. [20]

Curie-Weiss fitting is performed in a temperature range of 340 – 390 K (Fig. S1) using the *curve_fit* function in SciPy Python[31] according to the modified Curie-Weiss law[32]: $\chi = \frac{C}{T-\theta_{CW}} + \chi_0$. The fitting results are summarized in Table S4. For $Fe_{1.17}Sb$, the fit yields a Curie-Weiss temperature ($\theta_{CW}$) of 17.9K and an effective magnetic moment of 3.01 $\mu_B$/Fe. On the other hand, $Fe_{1.23}Sb$ results in a negative $\theta_{CW}$ of –24.3 K. The small value and sign switch of $\theta_{CW}$ suggest competing exchange interactions. The refined effective moment increases to 5.02 $\mu_B$/Fe, close to the spin-only values expected for high-spin $Fe^{2+}$ (4.90 $\mu_B$). Note that for $Fe_{1.30}Sb$, the Curie-Weiss fit yields unphysical parameters, possibly due to an additional magnetic transition above the detected temperature range.

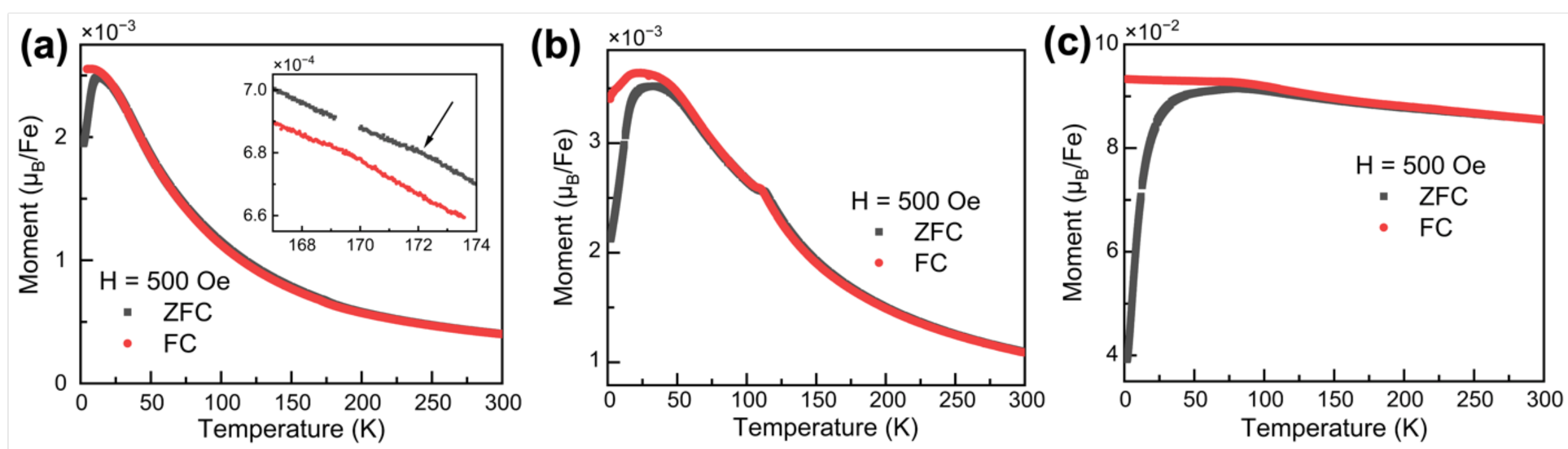

**Figure 3.** Magnetization vs. temperature under zero field cool (ZFC) and field cool (FC) for (a) $Fe_{1.17}Sb$, (b) $Fe_{1.23}Sb$, and (c) $Fe_{1.30}Sb$ powder samples.

To better identify the $T_N$'s in $Fe_{1+\delta}Sb$ and their magnetic structures, temperature-dependent neutron diffraction measurements were performed for $Fe_{1.17}Sb$ at NOMAD, SNS, ORNL, between 2 and 300 K. The corresponding time-of-flight (TOF) diffraction patterns are shown in Fig. 4a. Upon cooling, a prominent magnetic Bragg peak emerges at 150 but not at 200 K, indicating that $T_N$ lies between them. This indicates that the faint anomaly in magnetization (Fig. 3a) at ~170 K is the $T_N$. The magnetic peak at ~ 6.13 Å is $\sqrt{3}$ times of $d_{100}$ = 3.54 Å suggests the expansion of the magnetic superlattice. No additional magnetic Bragg peaks or discernible changes in the diffraction patterns are observed below 150 K and down to 2 K, suggesting that broad magnetic maximum and bifurcation near 20 K does not originate from a change in the long-range magnetic structure and originates from spin-glass freezing or defects. Low-temperature neutron diffraction measurements at 10 K using POWGEN at SNS of all three samples reveal the same magnetic peak positions (highlighted in Fig. 4b; full-range refinement is shown in Fig. S2) except with varied intensities.

The data can be indexed with a non-zero magnetic propagation vector $\boldsymbol{k}$ = (1/3, 1/3, 0). The magnetic structure refinement based on the magnetic space group $P\bar{6}'2'm$ yields a good fit for all three samples, shown in Fig. 4b.

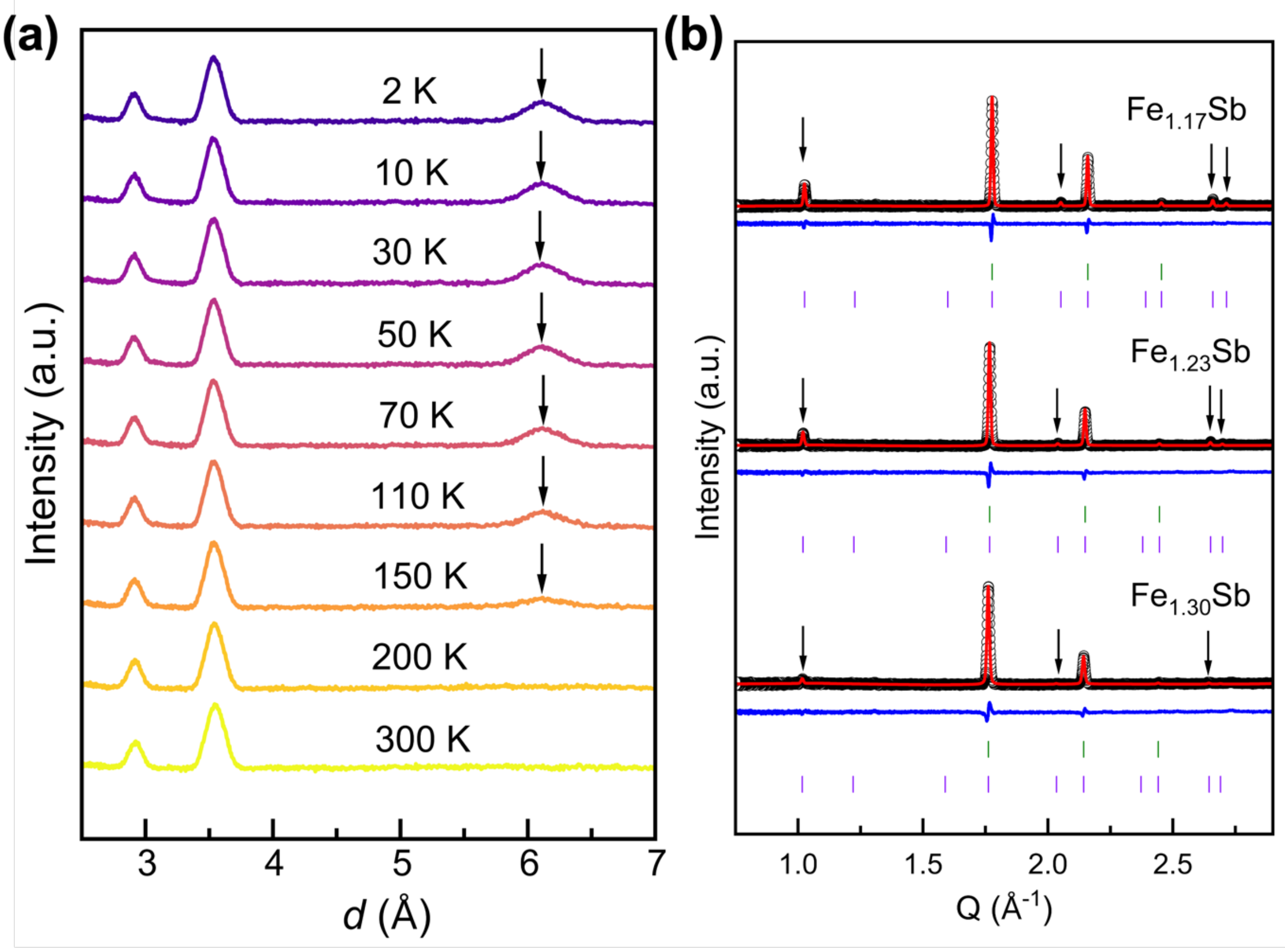

**Figure 4.** (a) Temperature-variant neutron diffraction of $Fe_{1.17}Sb$. The most prominent magnetic peaks are highlighted by black arrows. (b) Neutron diffraction refinement for $Fe_{1.17}Sb$, $Fe_{1.23}Sb$, and $Fe_{1.30}Sb$. Black open dots, red/blue curves, and green/purple ticks represent observation, calculated/difference and Bragg/magnetic peak positions, respectively.

The refined magnetic structure is shown in Fig. 5 and the corresponding refinement parameters are summarized in Table S5, featuring a coplanar compensated arrangement where Fe moments are rotated by 120° within the *ab* plane. The spin alignment remains parallel in adjacent layers along the *c*-axis, which agrees with a previous neutron study [17]. Additionally, increasing the interstitial Fe concentration from $Fe_{1.17}Sb$ to $Fe_{1.30}Sb$ monotonously reduces the refined Fe moment on the regular lattice site (2*a*) from 1.74 to 0.96 $\mu_B$/Fe. The refined moment for $Fe_{1.30}Sb$ carries a relatively large uncertainty of approximately 30%, due to its much reduced magnetic peak intensity (Fig. 4b). Previous studies have indicated that $Fe_{1+\delta}Sb$ exhibits metallic behavior while retaining localized magnetic moments. The suppressed local moments at higher Fe interstitial concentrations are likely due to the stronger Fe-Fe and Fe-Sb covalent interactions [33,34] especially along the *c* axis [34,35], which promote partial itineracy of the Fe 3$d$ electrons. The same mechanism likely also explains the observed $T_N$ lowering from $Fe_{1.17}Sb$ to $Fe_{1.23}Sb$.

Symmetry analysis of the refined magnetic structure indicates that it belongs to the class of non-relativistic spin-splitting antiferromagnet (NRSS AFMs), with altermagnets often considered the collinear subclass of this broader category. First, the translation-time reversal symmetry is broken. We use $\tau$ and $T$ to denote translational and time-reversal symmetry, respectively. The $\tau T$ symmetry is automatically broken in magnetic structures where $\boldsymbol{k}$ = (0, 0, 0) (the magnetic unit cell is of the same size as the atomic unit cell). In the case of $Fe_{1+\delta}Sb$, although $\boldsymbol{k}$ = (1/3, 1/3, 0) is non-zero and results in an expanded lattice in the ab plane of $\sqrt{3}a \times \sqrt{3}b$, the 120$^\circ$ triangular arrangement of the Fe moments dictates that any in-plane translational symmetry needs to be combined with a rotation symmetry, thereby breaking the $\tau T$ symmetry. In addition, the combined parity-time reversal ($PT$) symmetry is also broken, where $P$ denotes parity (inversion) symmetry. Because the Fe moments are aligned parallel across adjacent layers, no inversion-related pairs of antiparallel counterparts exist to preserve the $PT$ symmetry. The simultaneous breaking of both $\tau T$ and $PT$ in a compensated magnet gives rise to momentum-dependent spin splitting, identifying $Fe_{1+\delta}Sb$ as an altermagnet, whose polarized spin-momentum locking in the reciprocal space must exhibit the same symmetry as the crystal structure.

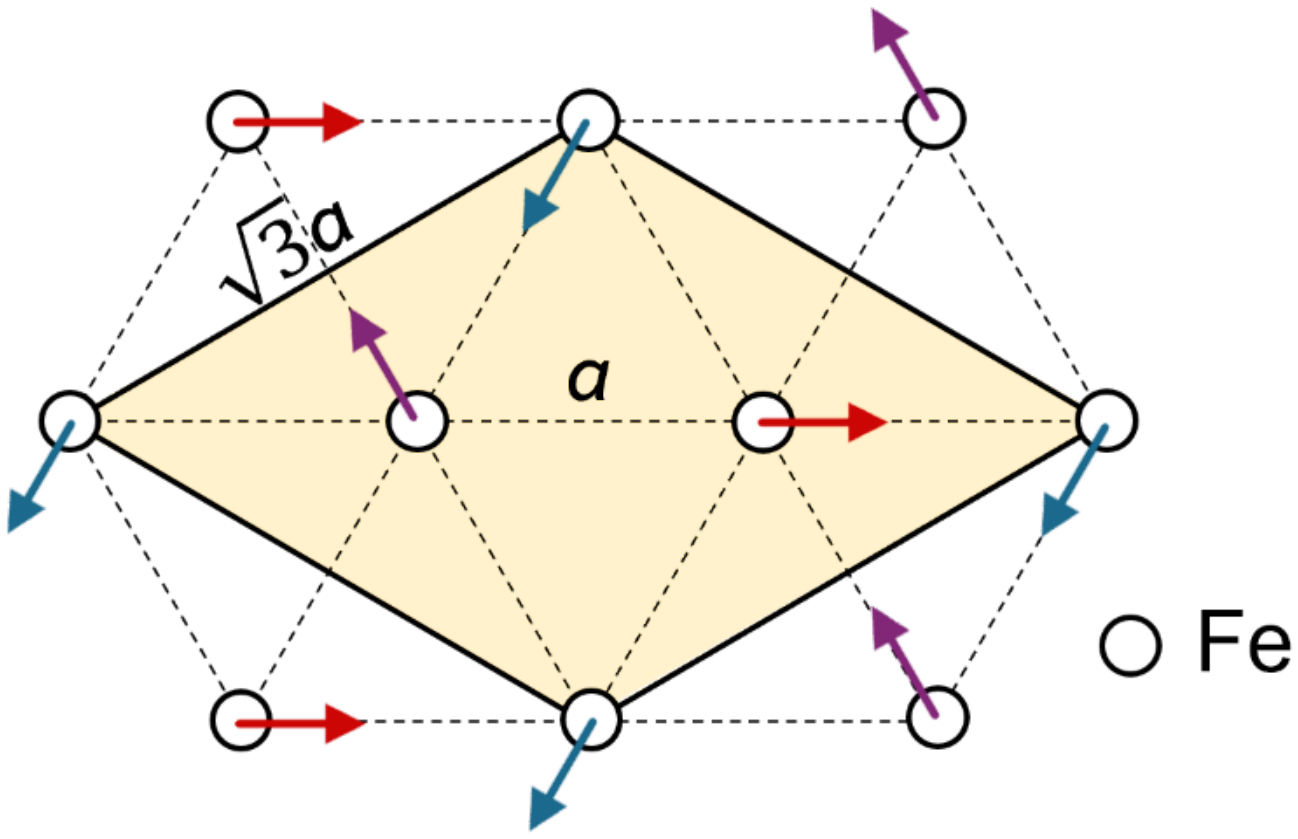

**Figure 5.** Magnetic structure of $Fe_{1+\delta}Sb$ in the layer of lattice Fe sites, as refined from neutron diffraction at 10 K. The $\sqrt{3} \times \sqrt{3}$ magnetic superlattice is highlighted in yellow with an expanded in-plane dimension of $\sqrt{3}a$. The moments align parallel between layers.

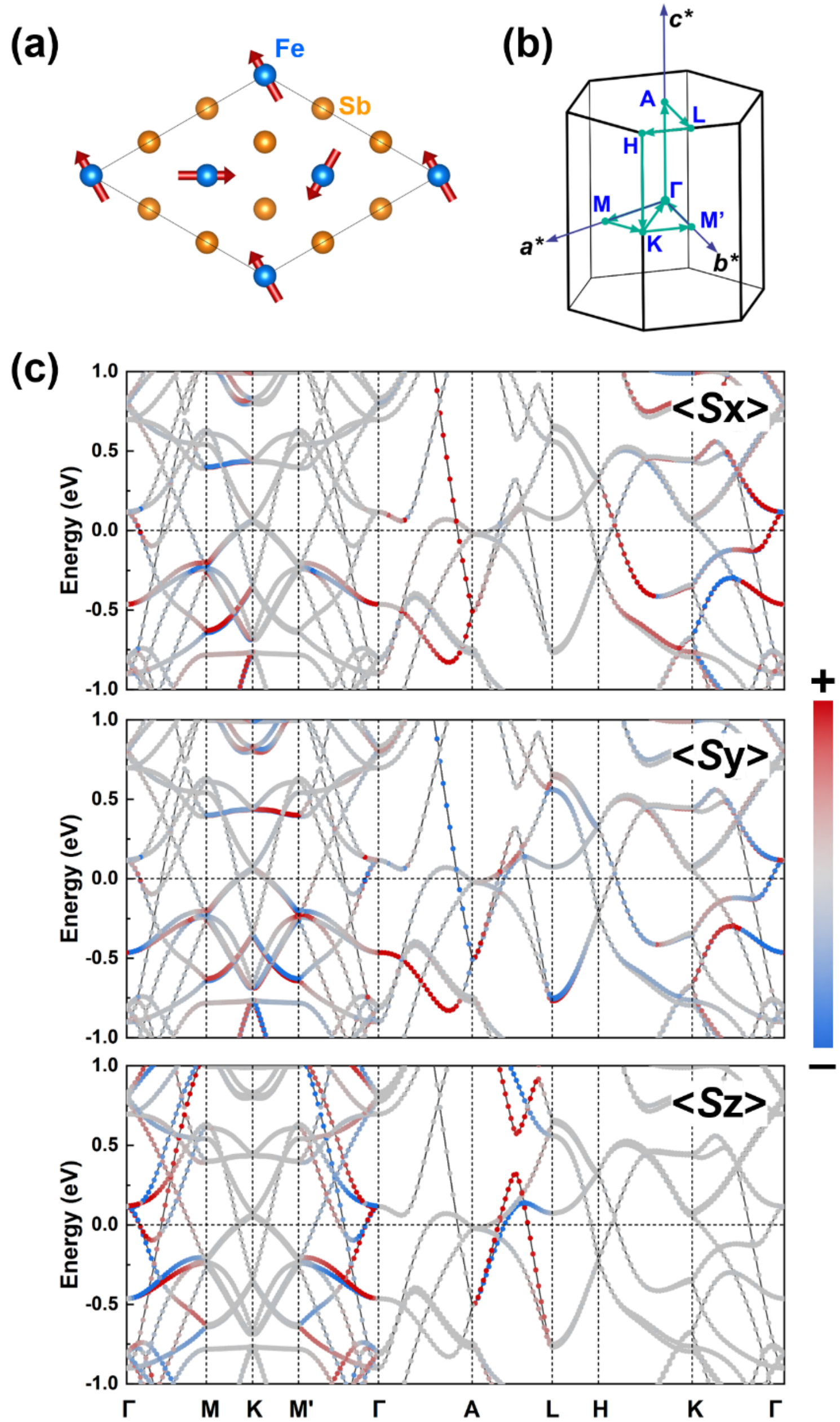

**Figure 6.** Altermagnetic spin splitting in FeSb based on a triangular coplanar magnetic order. (**a**) Coplanar magnetic configuration in the $\sqrt{3} \times \sqrt{3}$ superlattice. (**b**) Brillouin zone with the high-symmetry k-paths indicated. (c) DFT+U calculated band structures projected onto the $S_x$, $S_y$, and $S_z$ spin components. Red and blue indicate positive and negative projections, respectively, while gray denotes negligible polarization.

Based on the resolved $\sqrt{3} \times \sqrt{3}$ magnetic supercell as illustrated in Fig. 6a, we then calculate its band structure based on stoichiometric FeSb. In the absence of SOC, the calculated bands exhibit pronounced momentum-dependent spin splitting, e.g., along *Γ*–*M* and *Γ*–*M*′, as shown in Fig. 6c.

This behavior contrasts fundamentally with that of conventional ferromagnets, where the sign of the spin splitting remains uniform across the entire Brillouin zone. Furthermore, the spin polarization is dominated by the out-of-plane $S_z$ component, with only negligible contributions from the in-plane $S_x$ and $S_y$ components. Note that although finite $S_x$ and $S_y$ projections are present for some bands, the positive and negative spin components remain degenerate and overlap with each other, resulting in negligible in-plane spin splitting. The results indicate that the coplanar magnetic order generates an effective out-of-plane exchange field, leading to spin splitting even in the absence of out-of-plane magnetic moments. Furthermore, the $S_z$-projected bands remain spin-split throughout most of the Brillouin zone, except along reciprocal-space directions equivalent to *Γ*-*K*. The momentum-dependent spin polarization therefore forms a six-lobed pattern related by 3-fold rotation, as shown in Figs. 7, S3. This phenomenon can be rationalized within the framework of the recently developed spin-space-group classification of odd-parity altermagnetic systems.[36] The $\boldsymbol{k}$-dependent pattern is characteristic of the $f_{x^3-3xy^2}$-wave-like nonrelativistic spin splitting of odd-parity altermagnetism allowed for coplanar magnetic orders, with $s_z(k) = -s_z(-k)$ and $s_{x,y}(k) = 0$. Compared with the calculated band structures without SOC, the addition of SOC lifts several band degeneracies but has a negligible effect on the overall momentum-dependent spin-splitting characteristics, as shown in Fig. S4.

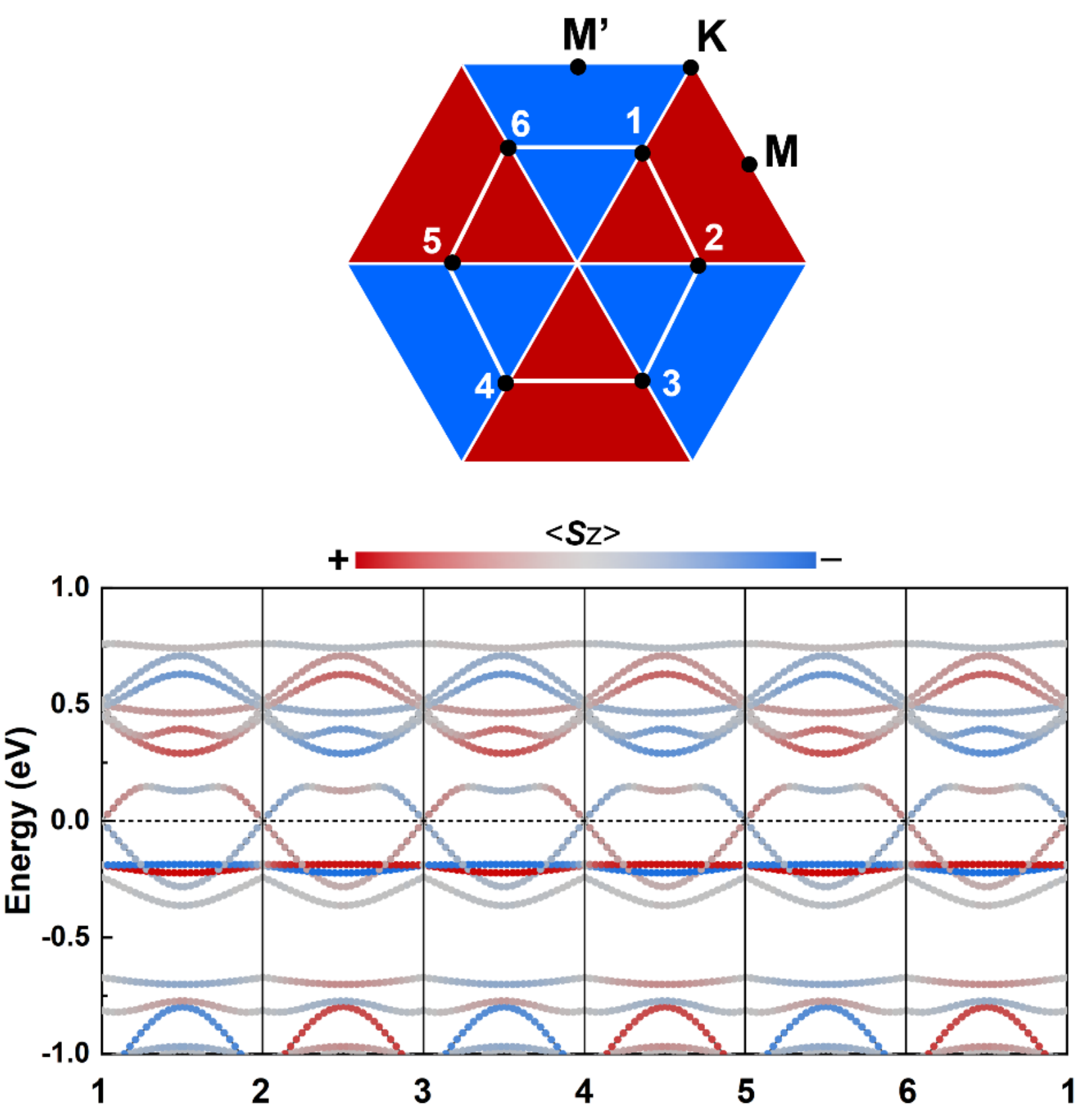

**Figure 7.** DFT+U calculated band structure of FeSb projected onto the out-of-plane spin component, *S*z, along the $\boldsymbol{k}$-paths 1-2-3-4-5-6 in the $k_z = 0$ plane. The corresponding paths in the hexagonal Brillouin zone

are illustrated in the upper panel. Red and blue indicate positive and negative projections, respectively, while gray denotes negligible polarization.

The fact that the iso-structural $Fe_{1+\delta}Sb$ and CrSb are both classified as altermagnets, with triangular coplanar and A-type out-of-plane collinear magnetic orders, respectively, motivates us to study the substituted series between the two in a nominal series $Fe_{1.2x}Cr_{1.2-1.2x}Sb$ ($x$ = 0, 0.2, 0.4, 0.6, 0.8, 1). A previous study found that the magnetic ground state of the $(Fe,Cr)_{1+\delta}Sb$ solid solution evolves from antiferromagnetic to ferromagnetic as the composition varies.[37] The possible rise of subtle ferromagnetism in altermagnets might provide a handle to tune the altermagnetic domains that are otherwise insusceptible to external magnetic fields. The x-ray diffraction (XRD) patterns of the as-synthesized $Fe_{1.2x}Cr_{1.2-1.2x}Sb$ series are shown in Fig. 8a and can all be indexed to the NiAs-type structure without additional peaks (except $Cr_{1.2}Sb$), indicating the formation of a continuous solid solution without a miscibility gap. Furthermore, the Bragg reflections systematically shift toward lower $d$-spacing with increasing Cr content, as shown in Fig. 8b, which is consistent with the trend reported [37] and indicates unit-cell volume expansion.

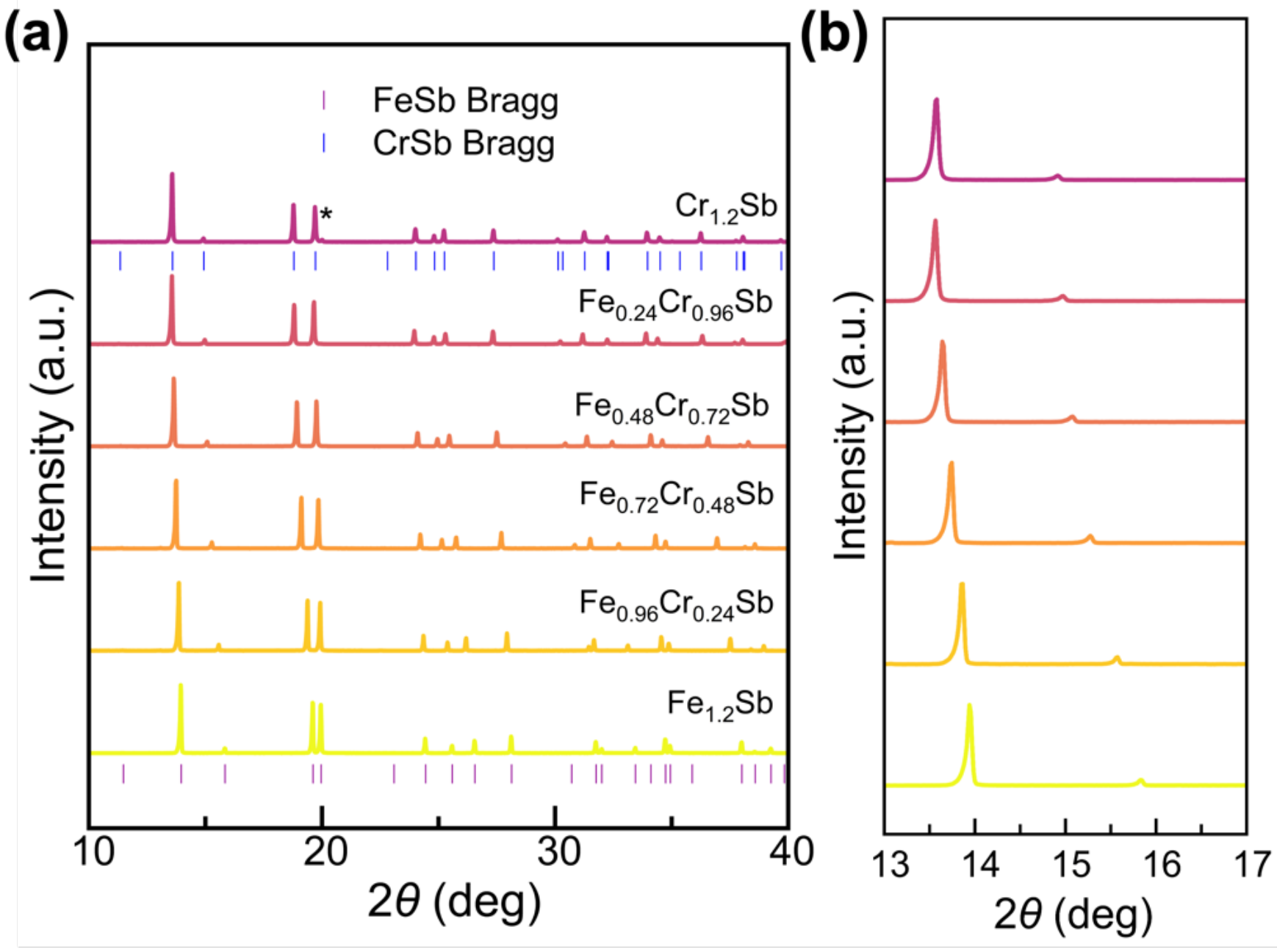

**Figure 8. (a)** Room temperature powder XRD patterns of the $Fe_{1.2x}Cr_{1.2-1.2x}Sb$ where $x$ = (0, 0.2, 0.4, 0.6, 0.8, 1) series. The expected Bragg positions based on the NiAs-type structures of pure $Fe_{1.2}Sb$ and $Cr_{1.2}Sb$ are indicated by purple and blue ticks, respectively. The asterisk (*) denotes an impurity peak. (b) Enlarged view of the 13˚- 17˚range.

The magnetic properties are investigated by DC magnetization measurements. As shown in Fig. 9a, $Fe_{0.96}Cr_{0.24}Sb$ and $Fe_{0.72}Cr_{0.48}Sb$ exhibit the largest magnetization among all compositions at low temperatures, consistent with the compositional dependence of the spontaneous ferromagnetic component $\mu_{Ferro}$ reported in the $(Fe,Cr)_{1+\delta}Sb$ system, where $\mu_{Ferro}$ reaches a maximum at 40-50% Cr substitution.[37] The precise origin of the enhanced $\mu_{Ferro}$ remains unresolved. One possible explanation is that partial Cr substitution introduces competing Fe-Fe, Fe-Cr, and Cr-Cr exchange interactions, thereby frustrating the establishment of a globally compensated magnetic order. A qualitatively similar trend has been reported in Fe-Cr alloys, where dilute Cr substitution enhances the Curie temperature through strong antiferromagnetic Fe-Cr nearest-neighbor exchange, which in turn reinforces the ferromagnetic ordering of the Fe moments.[38] The two compositions with the largest magnetization also possess the highest spin-freezing temperatures, suggesting that intermediate Cr substitution not only maximizes the uncompensated ferromagnetic component but also strengthens the magnetic interactions responsible for the emergence of spin-glass behavior. Furthermore, $Fe_{0.96}Cr_{0.24}Sb$ and $Fe_{0.72}Cr_{0.48}Sb$ undergo the second magnetic transition at higher temperatures (near 50 K) than the others. The low-temperature magnetic transition has been associated with spin-glass freezing in $Fe_{1+\delta}Sb$. Field-dependent magnetization measurements are performed at various temperatures for $Fe_{0.72}Cr_{0.48}Sb$, as shown Fig. 9b. The magnetization curves remain nonlinear up to 150 K, indicating the presence of a ferromagnetic component. A small hysteresis loop is observed up to 50 K (inset of Fig. 9b), which is close to the spin-glass transition temperature. Notably, the magnetization does not fully saturate even at an applied field of 9 T.

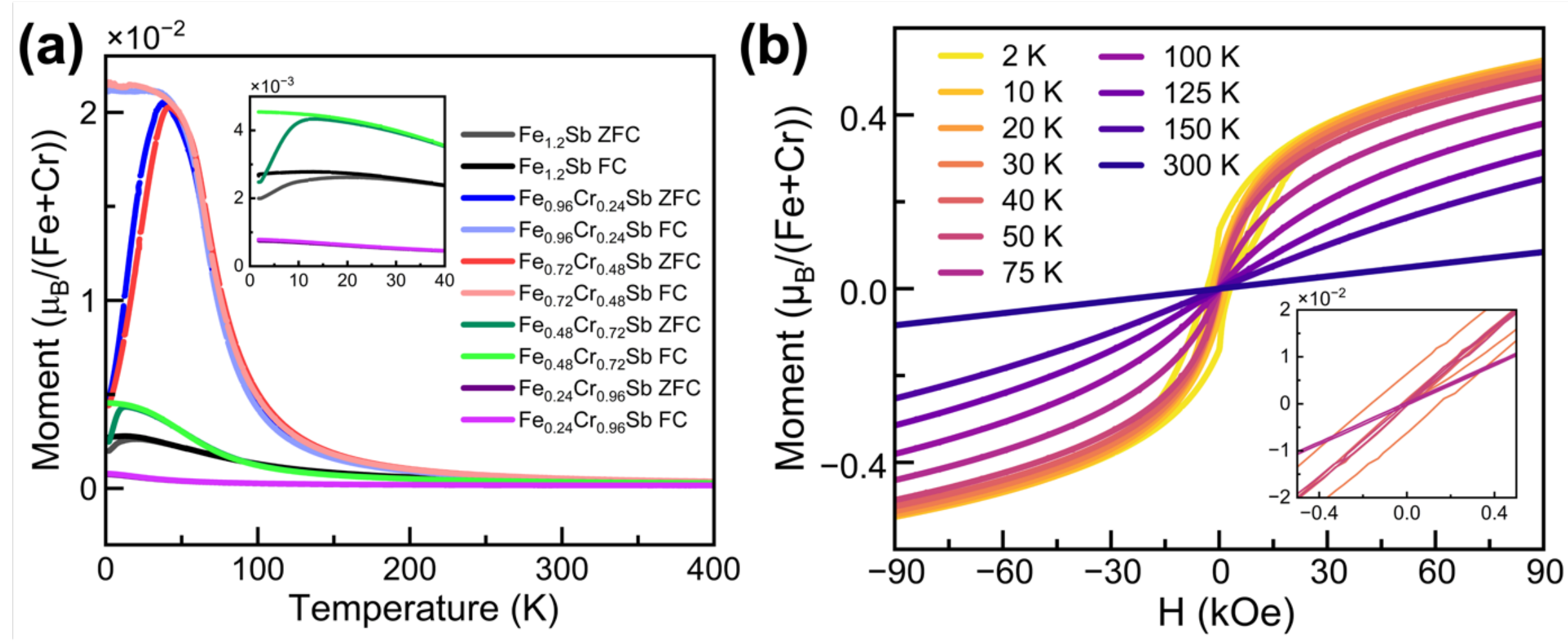

**Figure 9.** Magnetic properties of the $Fe_{1.2x}Cr_{1.2(1-x)}Sb$ (x = 0, 0.2, 0.4, 0.6, 0.8) series. (a) Temperature dependent ZFC and FC magnetization measured under an applied field of 500 Oe. (b) Isothermal M-H curves of $Fe_{0.72}Cr_{0.48}Sb$ measured at 2, 10, 20, 30, 40, 50, 75, 100, 125, 150, 300 K. The inset shows an enlarged view of the curves measured at 30, 40, 50, and 75 K.

The inferred spin-glass behavior is probed by AC susceptibility measurements on $Fe_{0.72}Cr_{0.48}Sb$. As shown in Fig. 10a, the real component of the AC susceptibility ($\chi$') exhibits a broad maximum near 56 K. A polynomial fit is applied to determine the freezing temperature ($T_F$) associated. We find that the extracted freezing temperatures systematically shift higher with increasing frequency (inset in Fig. 10a), from ~55.7 K at 100 Hz to ~56.8 K at 10 kHz, which is a hallmark of spin glasses. To further elucidate the nature of the spin-glass state, we apply the Mydosh parameter ($\Omega$) and the $E_a/(k_B T_0)$ ratio, which are proposed by Roy-Chowdhury et al. as robust criteria for distinguishing canonical and cluster spin glasses.[39] Canonical spin glasses arise from randomly distributed individual magnetic moments interacting through long-range oscillatory RKKY interactions,[40] while cluster spin glasses arise from short-range correlated magnetic clusters. The Mydosh parameter is defined as $\Omega = (T_{f2} - T_{f1})/T_{f1}(log f_2 - log f_1)$, where $f_1$ and $f_2$ are two excitation frequencies, and $T_{f1}$ and $T_{f2}$ are the corresponding freezing temperatures. The ratio $E_a/(k_B T_0)$ can be derive from the fitting the Vogel-Fulcher equation: $\tau = \tau_0 \exp\left[\frac{E_a}{k_B(T_f - T_0)}\right]$, where $\tau = 1/(2\pi f)$ is the characteristic relaxation time, $\tau_0$ is the relaxation time of individual spins or clusters, $T_f$ is the frequency-dependent freezing temperature obtained from the AC suspectibility data, $E_a$ is the activation energy, and $T_0$ is the Vogel-Fulcher temperature.[39] According to the criteria, canonical spin glasses satisfy $\Omega < 0.01$ and $E_a/(k_B T_0) < 1$, whereas cluster spin glasses exhibit $\Omega > 0.01$ and $E_a/(k_B T_0) > 1$.[39] For $Fe_{0.72}Cr_{0.48}Sb$, the obtained values of $\Omega = 0.01053$ and $E_a/(k_B T_0) = 2.89$ (derived from the Vogel-Fulcher fitting shown in Fig. 10b) indicate that the spin-glass state is best described as a cluster type.

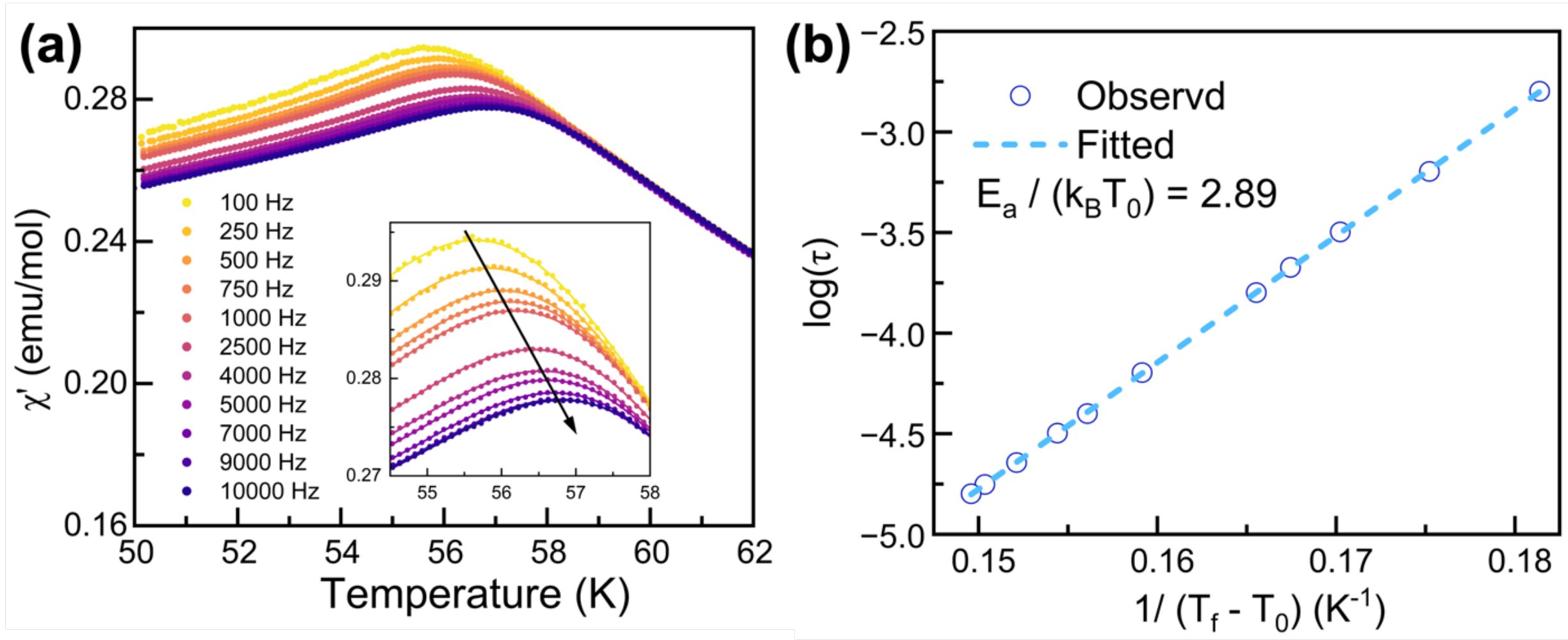

**Figure 10.** Dynamic magnetic behavior of $Fe_{0.72}Cr_{0.48}Sb$. (a) Temperature dependence of the real component of the AC magnetic susceptibility ($\chi$'). The inset enlarges the peak region. (b) Vogel-Fulcher analysis of the frequency-dependent freezing temperature.

## Acknowledgement

H.J. and C.-C.W. acknowledge the support from an NSF Career grant No. 2145832. A portion of this research used resources at the Spallation Neutron Source, a DOE Office of Science User Facility operated by the Oak Ridge National Laboratory.

## Conclusion

In summary, we establish $Fe_{1+\delta}Sb$ ($\delta = 0.17$-$0.30$) as a new non-collinear NRSS antiferromagnet in the NiAs family, by combining neutron diffraction, PDF analysis, magnetic measurements, and first principles calculations. Neutron diffraction reveals a compensated 120° coplanar magnetic order with $\boldsymbol{k} = (1/3, 1/3, 0)$ across the range of interstitial Fe concentrations investigated, while PDF refinements indicate that increasing Fe interstitial introduces local orthorhombic symmetry lowering. DFT calculations predict *f*-wave-like odd-parity momentum-dependent spin splitting that is dominated by the out-of-plane spin component despite the in-plane magnetic moments. We further investigate the $Fe_{1.2x}Cr_{1.2-1.2x}Sb$ solid solution and show that Cr substitution enhances the low-temperature magnetization and gives rise to a cluster spin-glass behavior at intermediate Cr concentrations through competing exchange interactions. These results establish $Fe_{1+\delta}Sb$ as a promising platform for exploring non-collinear NRSS antiferromagnetism, with metal interstitial and chemical substitution providing effective routes for tuning the magnetic order.

## REFERENCE

[1] L. Šmejkal, J. Sinova, and T. Jungwirth, Beyond Conventional Ferromagnetism and Antiferromagnetism: A Phase with Nonrelativistic Spin and Crystal Rotation Symmetry, Phys. Rev. X. **12**, 031042 (2022).

[2] L. Šmejkal, R. González-Hernández, T. Jungwirth, and J. Sinova, Crystal time-reversal symmetry breaking and spontaneous Hall effect in collinear antiferromagnets, Sci. Adv. **6**, eaaz8809.

[3] C. Song, H. Bai, Z. Zhou, L. Han, H. Reichlova, J. H. Dil, J. Liu, X. Chen, and F. Pan, Altermagnets as a new class of functional materials, Nat. Rev. Mater. **10**, 473 (2025).

[4] C.-C. Wei, E. Lawrence, A. Tran, and H. Ji, Crystal Chemistry and Design Principles of Altermagnets, ACS Org. Inorg. Au. **4**, 604 (2024).

[5] Z. Li *et al.*, Fully Field-Free Spin-Orbit Torque Switching Induced by Spin Splitting Effect in Altermagnetic RuO2, Adv. Mater. **37**, 2416712 (2025).

[6] Y. Guo *et al.*, Magnetic memory driven by spin splitting torque in nonrelativistic collinear antiferromagnet, Nat. Commun. **17**, 1309 (2025).

[7] B. Sekh, H. Rahaman, R. S. Verma, R. Maddu, K. Jawahar, and S. Piramanayagam, Spin splitting torque enabled artificial neuron with self-reset via synthetic antiferromagnetic coupling, arXiv preprint arXiv:2602.01874 (2026).

[8] Z. Feng *et al.*, An anomalous Hall effect in altermagnetic ruthenium dioxide, Nat. Electron. **5**, 735 (2022).

[9] T. Osumi, S. Souma, T. Aoyama, K. Yamauchi, A. Honma, K. Nakayama, T. Takahashi, K. Ohgushi, and T. Sato, Observation of a giant band splitting in altermagnetic MnTe, Phys. Rev. B. **109**, 115102 (2024).

[10] S. Reimers *et al.*, Direct observation of altermagnetic band splitting in CrSb thin films, Nat. Commun. **15**, 2116 (2024).

[11] E. A. Lawrence *et al.*, Fe Site Order and Magnetic Properties of $Fe_{1/4}NbS_2$, lnorg. Chem. **62**, 18179 (2023).

[12] C.-C. Wei *et al.*, $La_2O_3Mn_2Se_2$: A correlated insulating layered d-wave altermagnet, Phys. Rev. Mater. **9**, 024402 (2025).

[13] X. Zhang, J.-X. Xiong, L.-D. Yuan, and A. Zunger, Prototypes of Nonrelativistic Spin Splitting and Polarization in Symmetry Broken Antiferromagnets, Phys. Rev. X. **15**, 031076 (2025).

[14] S.-W. Cheong and F.-T. Huang, Altermagnetism classification, npj Quantum Mater. **10**, 38 (2025).

[15] M. Hu, O. Janson, C. Felser, P. McClarty, J. van den Brink, and M. G. Vergniory, Spin Hall and Edelstein effects in chiral non-collinear altermagnets, Nat. Commun. **16**, 8529 (2025).

[16] C.-C. Wei *et al.*, Symmetry-Protected Weyl Nodal Loops in a Triangular Altermagnet, arXiv preprint arXiv:2606.02527 (2026).

[17] T. Yashiro, Y. Yamaguchi, S. Tomiyoshi, N. Kazama, and H. Watanabe, Magnetic Structure of $Fe_{1+\delta}Sb$, J. Phys. Soc. Jpn. **34**, 58 (1973).

[18] A. Tiwari, P. D. Babu, M. Acet, K. R. Priolkar, and P. A. Bhobe, Effect of off-stoichiometry on structural, transport, and magnetic properties of altermagnetic CrSb, Phys. Rev. B. **113**, 144420 (2026).

[19] K. Yamaguchi, H. Yamamoto, Y. Yamaguchi, and H. Watanabe, Antiferromagnetism of $Fe_{1+\delta}Sb$, J. Phys. Soc. Jpn. **33**, 1292 (1972).

[20] P. J. Picone and P. E. Clark, Magnetic ordering of interstitial iron in $Fe_{1+x}Sb$ alloys, J. Magn. Magn. Mater. **25**, 140 (1981).

[21] B. H. Toby and R. B. Von Dreele, GSAS-II: the genesis of a modern open-source all purpose crystallography software package, J. Appl. Crystallogr. **46**, 544 (2013).

[22] C. L. Farrow, P. Juhas, J. W. Liu, D. Bryndin, E. S. Božin, J. Bloch, T. Proffen, and S. J. L. Billinge, PDFfit2 and PDFgui: computer programs for studying nanostructure in crystals, J. Phys. Condens. Matter. **19**, 335219 (2007).

[23] G. Kresse and J. Furthmüller, Efficiency of ab-initio total energy calculations for metals and semiconductors using a plane-wave basis set, Comput. Mater. Sci. **6**, 15 (1996).

[24] J. P. Perdew, K. Burke, and M. Ernzerhof, Generalized Gradient Approximation Made Simple, Phys. Rev. Lett. **77**, 3865 (1996).

[25] J. P. Perdew, K. Burke, and M. Ernzerhof, Generalized Gradient Approximation Made Simple [Phys. Rev. Lett. 77, 3865 (1996)], Phys. Rev. Lett. **78**, 1396 (1997).

[26] S. L. Dudarev, G. A. Botton, S. Y. Savrasov, C. J. Humphreys, and A. P. Sutton, Electron-energy-loss spectra and the structural stability of nickel oxide: An LSDA+U study, Phys. Rev. B. **57**, 1505 (1998).

[27] A. Rohrbach, J. Hafner, and G. Kresse, Ab initio study of the (0001) surfaces of hematite and chromia: Influence of strong electronic correlations, Phys. Rev. B. **70**, 125426 (2004).

[28] D. Hobbs, G. Kresse, and J. Hafner, Fully unconstrained noncollinear magnetism within the projector augmented-wave method, Phys. Rev. B. **62**, 11556 (2000).

[29] V. Wang, N. Xu, J.-C. Liu, G. Tang, and W.-T. Geng, VASPKIT: A user-friendly interface facilitating high-throughput computing and analysis using VASP code, Comput. Phys. Commun. **267**, 108033 (2021).

[30] G. Ren *et al.*, Atomic-Scale Observation of Symmetry Breaking in Altermagnetic MnTe, arXiv preprint arXiv:2605.27543 (2026).

[31] P. Virtanen *et al.*, SciPy 1.0: fundamental algorithms for scientific computing in Python, Nat. Methods **17**, 261 (2020).

[32] M. Sam and A. M. Hallas, Tutorial: a beginner's guide to interpreting magnetic susceptibility data with the Curie-Weiss law, Commun. Phys. **5** (2022).

[33] P. Amornpitoksuk, D. Ravot, A. Mauger, and J. C. Tedenac, Structural and magnetic properties of the ternary solid solution between CoSb and $Fe_{1+\delta}Sb$, Phys. Rev. B. **77**, 144405 (2008).

[34] R. Coehoorn, C. Haas, and R. A. de Groot, Electronic structure of MnSb, Phys. Rev. B. **31**, 1980 (1985).

[35] A. Kjekshus and W. B. Pearson, Phases with the nickel arsenide and closely-related structures, Prog. Solid State Chem. **1**, 83 (1964).

[36] X.-J. Luo, J.-X. Hu, M.-L. Hu, and K. Law, Spin Group Symmetry Criteria for Odd-parity Magnets, arXiv preprint arXiv:2510.05512 (2025).

[37] K. Yamaguchi, H. Watanabe, H. Yamamoto, and Y. Yamaguchi, Magnetic Properties of the Systems (Cr, Fe) Sb and (Cr, Co) Sb, J. Phys. Soc. Jpn. **31**, 1042 (1971).

[38] J. B. Chapman, P.-W. Ma, and S. L. Dudarev, Dynamics of magnetism in Fe-Cr alloys with Cr clustering, Phys. Rev. B. **99**, 184413 (2019).

[39] M. Roy-Chowdhury, M. S. Seehra, and S. Thota, Optimized analysis of the AC magnetic susceptibility data in several spin-glass systems using the Vogel–Fulcher and Power laws, AIP Adv. **13** (2023).

[40] H. Kawamura and T. Taniguchi, in *Handbook of Magnetic Materials*, edited by K. H. J. Buschow (Elsevier, 2015), pp. 1.

**Supplementary Information for**
**Non-relativistic spin splitting in a triangular metal-excess magnet $Fe_{1+\delta}Sb$**

[1]Chao-Chun Wei, [2]Xiaojuan Ni, [1,3]Sophia Adams, [4]Jacob Kjeldahl Jensen, [5]Jue Liu, [5]Qiang Zhang, [4]Luisa Whittaker-Brooks, [1]Huiwen Ji*

[1]*Department of Materials Science and Engineering, University of Utah, Salt Lake City, Utah 84112, United States*

[2]*Department of Chemistry and Biochemistry, The University of Arizona, Tucson, Arizona 85721-0041, United States*

[3]*Department of Chemical and Biomolecular Engineering, Johns Hopkins University, Baltimore, Maryland, 21218, United States*

[4]*Department of Chemistry, University of Utah, Salt Lake City, Utah 84112, United States*

[5]*Neutron Sciences Division, Oak Ridge National Laboratory, Oak Ridge, Tennessee 37831, United States*

Corresponding: huiwen.ji@utah.edu

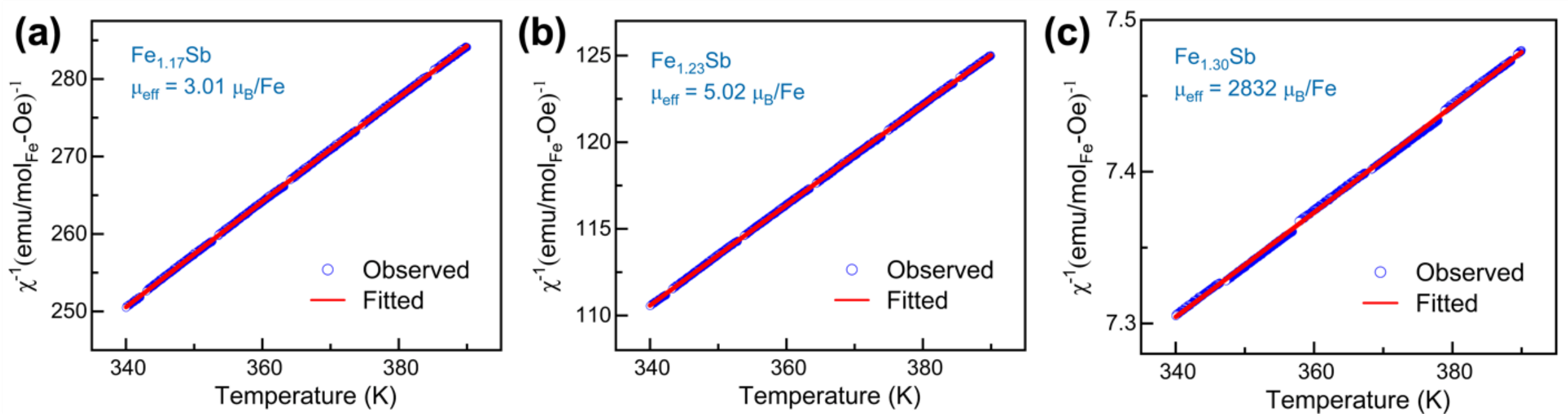

**Figure S1.** Inverse magnetic susceptibility ($\chi^{-1}$) as a function of temperature for (a) $Fe_{1.17}Sb$, (b) $Fe_{1.23}Sb$, and (c) $Fe_{1.30}Sb$ under the external magnetic field of 5000 Oe.

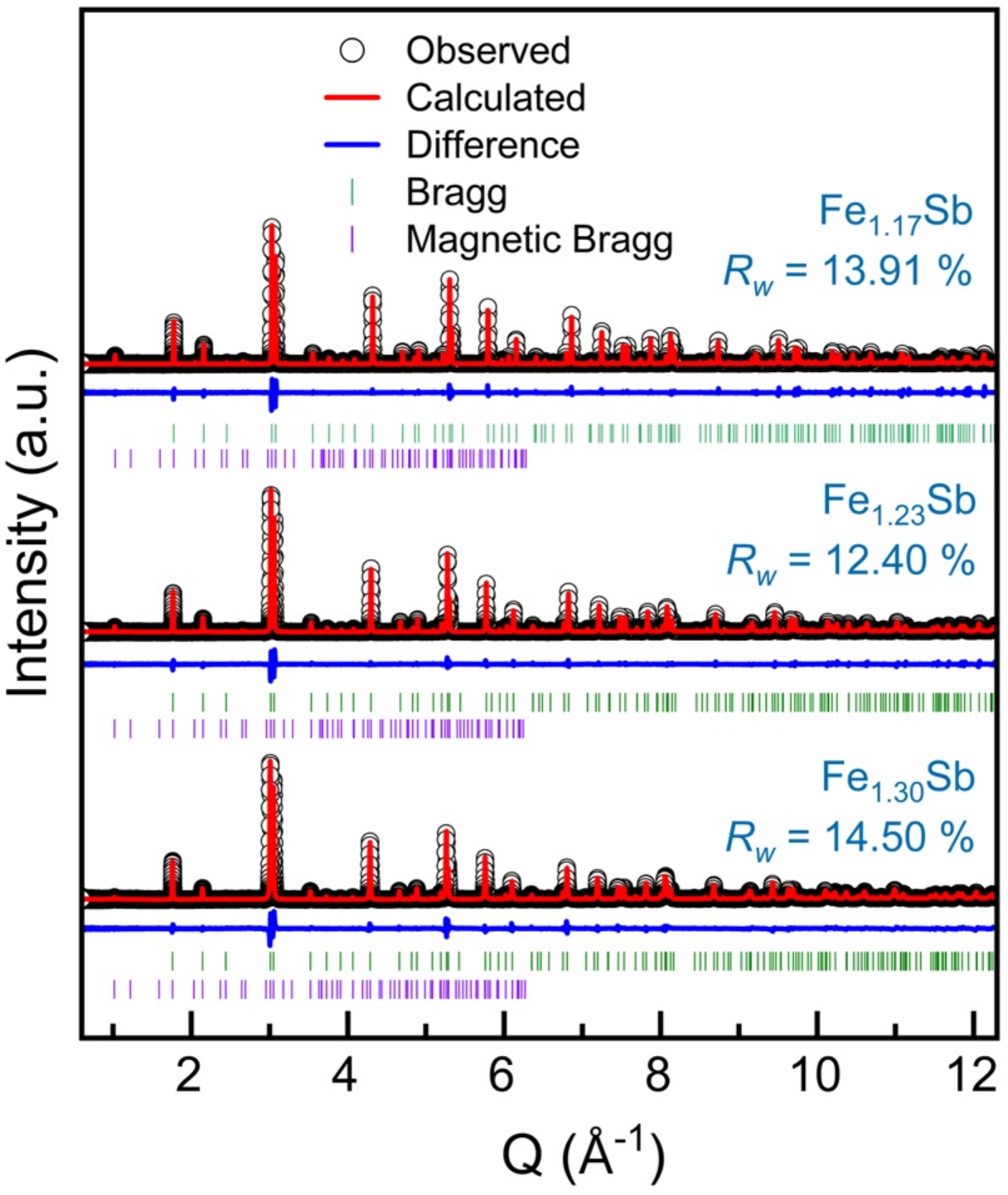

**Figure S2.** Rietveld refinements of the POWGEN neutron powder diffraction patterns for $Fe_{1+\delta}Sb$ collected at 10 K over the full Q range of 0.6 to 12.3 $Å^{-1}$.

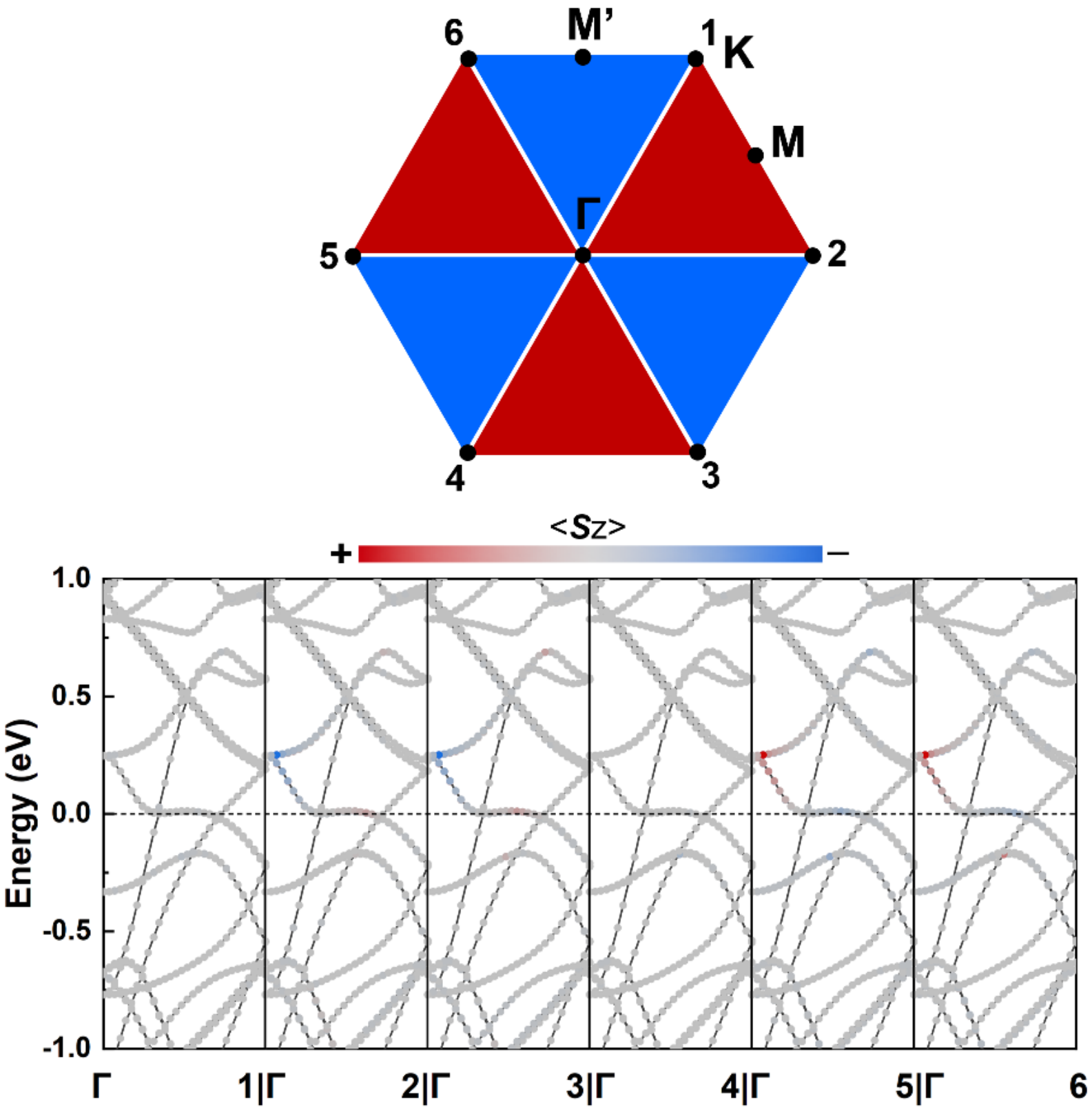

**Figure S3.** DFT+U calculated band structure of FeSb projected onto the out-of-plane spin component, $S_z$, along the ***k***-paths $\Gamma$–1, $\Gamma$–2, $\Gamma$–3, $\Gamma$–4, $\Gamma$–5, and $\Gamma$–6. The corresponding paths in the hexagonal Brillouin zone are illustrated in the upper panel. Red and blue indicate positive and negative projections, respectively, while gray denotes negligible polarization.

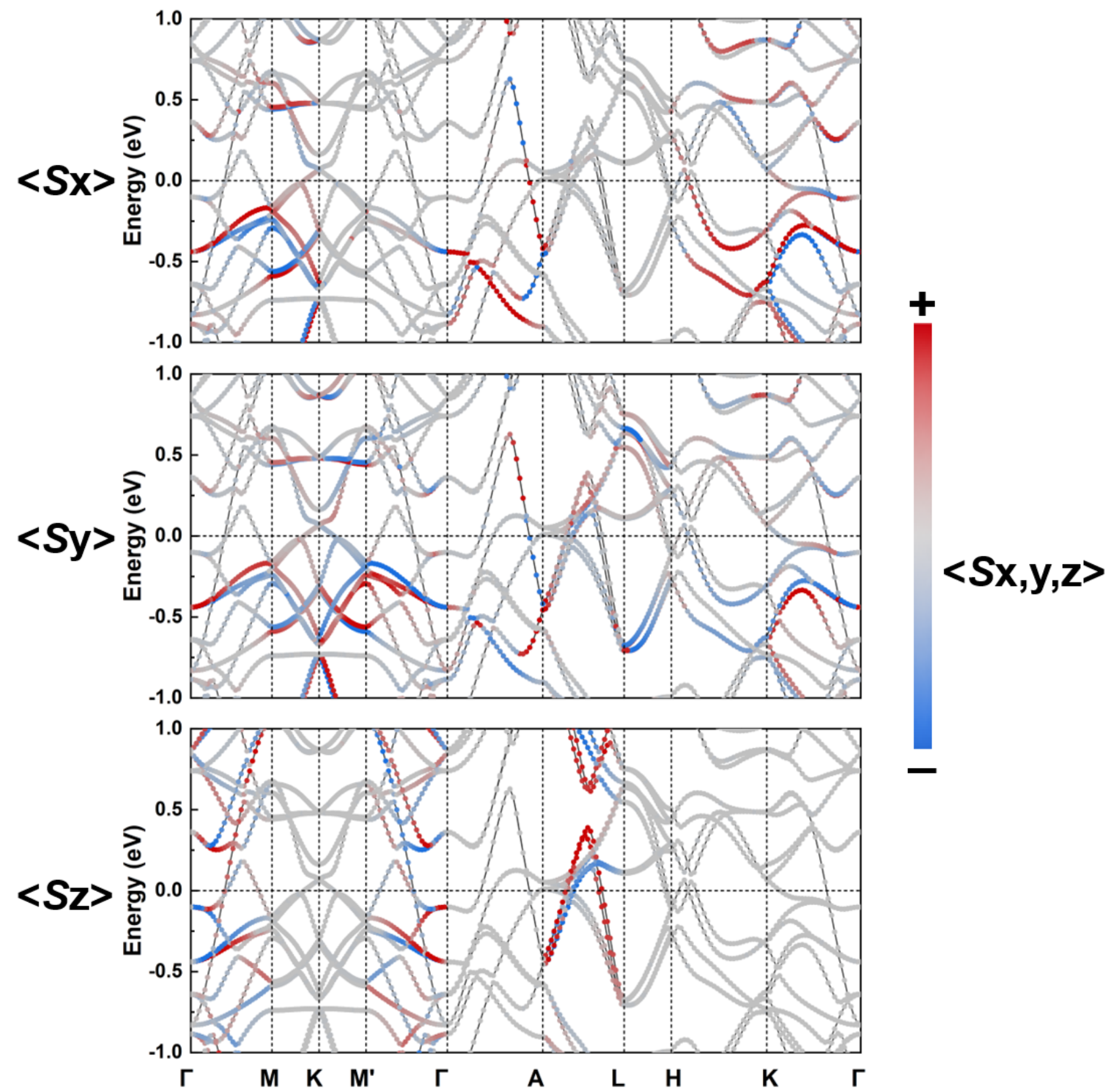

**Figure S4.** DFT+U calculated band structures projected onto the $S_x$, $S_y$, and $S_z$ spin components with consideration of SOC. Red and blue indicate positive and negative projections, respectively, while gray denotes negligible polarization.

**Table S1.** Refinement table for NOMAD data Neutron Diffraction refinement.

| **Chemical Formula (Nominal)** | **Space Group** | ***a*** **(Å)** | ***c*** **(Å)** | ***V*****(Å$^3$)** | ***R*****$_w$(%)** |
|---|---|---|---|---|---|
| $Fe_{1.2}Sb$ | $P6_3/mmc$ (Hexagonal) | 4.09196(4) | 5.14486(4) | 73.6048(10) | 4.90 |
| $Fe_{1.3}Sb$ | | 4.11660(6) | 5.16418(6) | 75.7897(14) | 9.31 |
| $Fe_{1.4}Sb$ | | 4.12553(7) | 5.17266(7) | 76.2437(15) | 4.70 |

| $Fe_{1.2}Sb$ Atomic Refinement | | | | | | |
|---|---|---|---|---|---|---|
| **Atom** | **Site** | ***x*** | ***y*** | ***z*** | ***U*****$_{iso}$ (Å$^2$)** | **Occupancy** |
| Fe1 | 2a | 0 | 0 | 0 | 0.00755(8) | 1 |
| Sb | 2c | 0.33333 | 0.66667 | 0.25 | 0.01142(18) | 1 |
| Fe2 | 2d | 0.66667 | 0.33333 | 0.25 | 0.0078(7) | 0.1748(13) |
| | | | | | | |
| $Fe_{1.3}Sb$ Atomic Refinement | | | | | | |
| **Atom** | **Site** | ***x*** | ***y*** | ***z*** | ***U*****$_{iso}$ (Å$^2$)** | **Occupancy** |
| Fe1 | 2a | 0 | 0 | 0 | 0.00932(10) | 1 |
| Sb | 2c | 0.33333 | 0.66667 | 0.25 | 0.01590(27) | 1 |
| Fe2 | 2d | 0.66667 | 0.33333 | 0.25 | 0.0091(7) | 0.2295(18) |
| | | | | | | |
| $Fe_{1.4}Sb$ Atomic Refinement | | | | | | |
| **Atom** | **Site** | ***x*** | ***y*** | ***z*** | ***U*****$_{iso}$ (Å$^2$)** | **Occupancy** |
| Fe1 | 2a | 0 | 0 | 0 | 0.00607(9) | 1 |
| Sb | 2c | 0.33333 | 0.66667 | 0.25 | 0.01239(30) | 1 |
| Fe2 | 2d | 0.66667 | 0.33333 | 0.25 | 0.0099(6) | 0.3018(19) |

**Table S2.** PDFgui refinement of the short-range atomic structure (1.5–10 Å) for $Fe_{1+\delta}Sb$ using the hexagonal NiAs-type structural model (space group $P6_3/mmc$) based on NOMAD neutron total scattering data at 300 K.

| **Chemical Formula** | ***a*** **(Å)** | ***c*** **(Å)** | **Scale Factor** | **Delta1** | $R_w$**(%)** |
|---|---|---|---|---|---|
| $Fe_{1.17}Sb$ | 4.0936(33) | 5.1497(77) | 0.776(23) | 1.76(12) | 7.077 |
| $Fe_{1.23}Sb$ | 4.1194(24) | 5.1721(58) | 1.177(25) | 1.634(93) | 7.082 |
| $Fe_{1.30}Sb$ | 4.1215(37) | 5.1769(91) | 0.666(23) | 1.63(16) | 9.980 |

| $Fe_{1.17}Sb$ Atomic Refinement | | | | | | |
|---|---|---|---|---|---|---|
| **Atom** | **Site** | ***x*** | ***y*** | ***z*** | $U_{iso}$ **(Å²)** | **Occupancy** |
| Fe1 | 2a | 0 | 0 | 0 | 0.01147(87) | 1 |
| Sb | 2c | 0.33333 | 0.66667 | 0.25 | 0.0211(54) | 1 |
| Fe2 | 2d | 0.66667 | 0.33333 | 0.25 | 0.010(9) | 0.178(46) |
| | | | | | | |
| $Fe_{1.23}Sb$ Atomic Refinement | | | | | | |
| **Atom** | **Site** | ***x*** | ***y*** | ***z*** | $U_{iso}$ **(Å²)** | **Occupancy** |
| Fe1 | 2a | 0 | 0 | 0 | 0.01253(69) | 1 |
| Sb | 2c | 0.33333 | 0.66667 | 0.25 | 0.0248(54) | 1 |
| Fe2 | 2d | 0.66667 | 0.33333 | 0.25 | 0.0139(7) | 0.244(40) |
| | | | | | | |
| $Fe_{1.30}Sb$ Atomic Refinement | | | | | | |
| **Atom** | **Site** | ***x*** | ***y*** | ***z*** | $U_{iso}$ **(Å²)** | **Occupancy** |
| Fe1 | 2a | 0 | 0 | 0 | 0.01003(93) | 1 |
| Sb | 2c | 0.33333 | 0.66667 | 0.25 | 0.0244(84) | 1 |
| Fe2 | 2d | 0.66667 | 0.33333 | 0.25 | 0.0131(99) | 0.243(65) |

**Table S3.** PDFgui refinement results of the short-range atomic structure (1.5–10 Å) for $Fe_{1+\delta}Sb$ using the orthorhombic *Amm*2 structural model fitted to NOMAD neutron total scattering data at 300 K.

| Chemical Formula | *a* (Å) | *b* (Å) | *c* (Å) | Scale Factor | Delta1 | $R_w$(%) |
|---|---|---|---|---|---|---|
| $Fe_{1.17}Sb$ | 5.1478(82) | 4.092(14) | 7.101(26) | 0.766(62) | 1.80(15) | 6.052 |
| $Fe_{1.23}Sb$ | 5.1697(61) | 4.1158(92) | 7.148(19) | 1.138(33) | 1.63(13) | 4.999 |
| $Fe_{1.30}Sb$ | 5.1743(94) | 4.125(11) | 7.140(24) | 0.632(34) | 1.72(19) | 6.236 |

| $Fe_{1.17}Sb$ Atomic Refinement | | | | | | |
|---|---|---|---|---|---|---|
| **Atom** | **Site** | ***x*** | ***y*** | ***z*** | $U_{iso}$ (Å$^2$) | **Occupancy** |
| Fe1 | 4c | 0.75 | 0 | 0 | 0.0155(37) | 0.98(11) |
| Sb1 | 2a | 0 | 0 | 0.661(13) | 0.009(13) | 1 |
| Sb2 | 2b | 0.5 | 0 | 0.351(12) | 0.028(16) | 1 |
| Fe2 | 2b | 0 | 0 | 0.6494(98) | 0.006(12) | 0.36(17) |
| Fe3 | 2a | 0 | 0 | 0.331(48) | 0.006(12) | 0.08(11) |
| $Fe_{1.23}Sb$ Atomic Refinement | | | | | | |
| **Atom** | **Site** | ***x*** | ***y*** | ***z*** | $U_{iso}$ (Å$^2$) | **Occupancy** |
| Fe1 | 4c | 0.75 | 0 | 0 | 0.0162(29) | 1 |
| Sb1 | 2a | 0 | 0 | 0.656(12) | 0.015(14) | 1 |
| Sb2 | 2b | 0.5 | 0 | 0.341(14) | 0.026(12) | 1 |
| Fe2 | 2b | 0 | 0 | 0.654(12) | 0.09(12) | 0.46(15) |
| Fe3 | 2a | 0 | 0 | 0.3762(89) | 0.004(14) | 0.1(1) |
| $Fe_{1.30}Sb$ Atomic Refinement | | | | | | |
| **Atom** | **Site** | ***x*** | ***y*** | ***z*** | $U_{iso}$ (Å$^2$) | **Occupancy** |
| Fe1 | 4c | 0.75 | 0 | 0 | 0.0128(47) | 1 |
| Sb1 | 2a | 0 | 0 | 0.651(12) | 0.014(27) | 1 |
| Sb2 | 2b | 0.5 | 0 | 0.339(18) | 0.02(2) | 1 |
| Fe2 | 2b | 0 | 0 | 0.655(15) | 0.01(2) | 0.59(28) |
| Fe3 | 2a | 0 | 0 | 0.38(2) | 0.01(2) | 0.13(13) |

**Table S4.** Curie-Weiss fitting parameters for $Fe_{1+\delta}Sb$ obtained by fitting the magnetic susceptibility measured between 340 and 390 K to Curie-Weiss law. The uncertainties represent one standard deviation (1σ) from the least-squares fit.

| **Materials** | $\chi_0$ **(emu/mol$_{Fe}$)** | $C$ **(emu*K/mol$_{Fe}$)** | $\theta_{CW}$ **(K)** | $\mu_{eff}$ **($\mu_B$/Fe)** |
|---|---|---|---|---|
| $Fe_{1.17}Sb$ | $(4.68 \pm 0.16) \times 10^{-4}$ | 1.135 ± 0.011 | 17.9 ± 1.7 | 3.01 ± 0.02 |
| $Fe_{1.23}Sb$ | $(3.84 \pm 0.46) \times 10^{-4}$ | 3.155 ± 0.036 | –24.3 ± 2.2 | 5.02 ± 0.03 |
| $Fe_{1.30}Sb$ | –7.872 ± 0.003 | 1003319.13± 2.46 | –124942.1 ± 39.8 | 2832.689 ± 0.004 |

**Table S5.** Refined structural and magnetic parameters obtained from Rietveld refinement of the POWGEN neutron powder diffraction data.

| **Chemical Formula** | ***a*** **(Å)** | ***c*** **(Å)** | ***V*(Å$^3$)** | **$M_x$ (1/2$M_y$)** | **$R_w$(%)** |
|---|---|---|---|---|---|
| $Fe_{1.17}Sb$ | 4.082623 (27) | 5.119761(24) | 73.9025(6) | 0.78(5) | 13.91 |
| $Fe_{1.23}Sb$ | 4.10639(3) | 5.135984(29) | 75.0025(7) | 0.55(5) | 12.40 |
| $Fe_{1.30}Sb$ | 4.11844(6) | 5.14679(5) | 75.6020(12) | 0.43(13) | 14.50 |

| $Fe_{1.17}Sb$ Atomic Refinement | | | | | | |
|---|---|---|---|---|---|---|
| **Atom** | **Site** | ***x*** | ***y*** | ***z*** | **$U_{iso}$ (Å$^2$)** | **Occupancy** |
| Fe1 | 2a | 0 | 0 | 0 | 0.00276(9) | 1 |
| Sb | 2c | 0.33333 | 0.66667 | 0.25 | 0.00742(16) | 1 |
| Fe2 | 2d | 0.66667 | 0.33333 | 0.25 | 0.0001 | 0.1623(16) |
| | | | | | | |
| $Fe_{1.23}Sb$ Atomic Refinement | | | | | | |
| **Atom** | **Site** | ***x*** | ***y*** | ***z*** | **$U_{iso}$ (Å$^2$)** | **Occupancy** |
| Fe1 | 2a | 0 | 0 | 0 | 0.00265(8) | 1 |
| Sb | 2c | 0.33333 | 0.66667 | 0.25 | 0.00959(19) | 1 |
| Fe2 | 2d | 0.66667 | 0.33333 | 0.25 | 0.0001 | 0.2210(16) |
| | | | | | | |
| $Fe_{1.30}Sb$ Atomic Refinement | | | | | | |
| **Atom** | **Site** | ***x*** | ***y*** | ***z*** | **$U_{iso}$ (Å$^2$)** | **Occupancy** |
| Fe1 | 2a | 0 | 0 | 0 | 0.00251(11) | 1 |
| Sb | 2c | 0.33333 | 0.66667 | 0.25 | 0.00966(29) | 1 |
| Fe2 | 2d | 0.66667 | 0.33333 | 0.25 | 0.0001 | 0.2671(25) |